\documentclass[journal, 12pt, letterpaper, draftclsnofoot, onecolumn]{IEEEtran}

\usepackage{amsfonts,cite,graphicx}
\usepackage[cmex10]{amsmath}
\newlength{\plotwidth}
\newtheorem{theorem}{Theorem}

\begin{document}
\title{Linear Processing and Sum Throughput in the Multiuser MIMO Downlink}
\author{
Adam~J.~Tenenbaum,~\IEEEmembership{Student~Member,~IEEE,} and~Raviraj~S.~Adve,~\IEEEmembership{Senior~Member,~IEEE}%
\thanks{\copyright 2009 IEEE. Personal use of this material is permitted. Permission from IEEE must be obtained for all other uses, including reprinting/republishing this material for advertising or promotional 
purposes, creating new collective works for resale or redistribution to servers or lists, or reuse of 
any copyrighted component of this work in other works.}%
\thanks{This manuscript has been accepted for publication in IEEE Transactions on Wireless Communications.}%
}
\maketitle
\begin{abstract}
We consider linear precoding and decoding in the downlink of a
multiuser multiple-input, multiple-output (MIMO) system, wherein each
user may receive more than one data stream.  We propose several mean
squared error (MSE) based criteria for joint transmit-receive
optimization and establish a series of relationships linking these
criteria to the signal-to-interference-plus-noise ratios of individual
data streams and the information theoretic channel capacity under
linear minimum MSE decoding.  In particular, we show that achieving
the maximum sum throughput is equivalent to minimizing the product of
MSE matrix determinants (PDetMSE).  Since the PDetMSE minimization
problem does not admit a computationally efficient solution, a
simplified scalar version of the problem is considered that minimizes
the product of mean squared errors (PMSE). An iterative algorithm is
proposed to solve the PMSE problem, and is shown to provide
near-optimal performance with greatly reduced computational
complexity.  Our simulations compare the achievable sum rates under
linear precoding strategies to the sum capacity for the broadcast
channel.
\end{abstract}

\section{Introduction}
The benefits of using multiple antennas for wireless communication
systems are well known.  When antenna arrays are present at the
transmitter and/or receiver, multiple-input multiple-output (MIMO)
techniques can utilize the spatial dimension to yield improved
reliability, increased data rates, and the spatial separation of
users.  In this paper, the methods we propose will focus on exploiting
all of these features, with the goal of maximizing the sum data rate
achieved in the MIMO multiuser downlink.

The optimal strategy for maximizing sum rate in the multiuser MIMO
downlink, also known as the broadcast channel (BC), was first proposed
in~\cite{CS03}; the authors prove that Costa's dirty paper
coding (DPC) strategy~\cite{Costa83} is sum capacity achieving for a
pair of single-antenna users.  The sum-rate optimality of DPC was
generalized to an arbitrary number of multi-antenna receivers using
the notions of game theory~\cite{YC04} and uplink-downlink
duality~\cite{VJG03,VT03}; this duality is employed in
\cite{SBJ03,JRVJG05} to derive iterative solutions that find the sum
capacity.  DPC has been shown to be the optimal precoding strategy not
only for sum capacity, but also for the entire capacity region in the
BC~\cite{WSS06}.  Unfortunately, finding a practical realization of the
DPC precoding strategy has proven to be a difficult problem.  Existing
solutions, which are largely based on Tomlinson-Harashima precoding
(THP)~\cite{WFVH04,LK05,FYL07,YVC05}, incur high complexity due to
their nonlinear nature and the combinatorial problem of user order
selection.  THP-based schemes also suffer from rate loss when compared
to the sum capacity due to modulo and shaping losses.

Linear precoding provides an alternative approach for transmission in
the MIMO downlink, trading off a reduction in precoder complexity for
suboptimal performance.  Orthogonalization based schemes use zero
forcing (ZF) and block diagonalization (BD) to transform the multiuser
downlink into parallel single-user systems~\cite{BK02,SS04}.  A
waterfilling power allocation can then be used to allocate powers to
each of the users~\cite{LJ06}.  The simplicity of these approaches
comes at the expense of an antenna constraint requiring at least as
many transmit antennas as the total number of receive antennas.  These
schemes, therefore, restrict the possibility of gains from additional
receiver antennas.  The constraint is relaxed under successive zero
forcing~\cite{DL07}, which requires only partial orthogonality but
incurs higher complexity in finding an optimal user ordering.
Coordinated beamforming~\cite{FSS03} and generalized
orthogonalization~\cite{PWN04} are able to avoid the antenna
constraint via iterative optimization of transmit and receive
beamformers.

It is also possible to improve the sum rate achieved with ZF and BD by
including user or antenna selection in the precoder design.  The
sum-rate maximizing ZF precoder can be found by comparing precoders for
all possible subsets of available receive antennas~\cite{CS03};
however, this strategy incurs exponential complexity on the order of
the total number of receive antennas.  Greedy and suboptimal strategies
for user selection~\cite{DS05, YG06, SCAHE06, WLZ08} may also be
applied with lower computational cost. However, user selection is
outside the scope of this paper; our goal here is to focus on the rates
achievable under linear precoding.  While all of these schemes possess lower complexity than
the THP based methods, the use of orthogonalization results in
suboptimal performance due to noise enhancement.  In this paper, we
consider the optimal formulation for sum rate maximization under
linear precoding.

Much of the existing literature on linear precoding for multiuser MIMO
systems focuses on minimizing the sum of mean squared errors (SMSE)
between the transmitted and received signals under a sum power
constraint~\cite{TA04,SB04,SS05,SSJB05,KTA06,MJHU06}.  An important
recurring theme in most of these papers is the use of an
uplink-downlink duality for both MSE and
signal-to-interference-plus-noise ratio (SINR) introduced
in~\cite{SB04} for the single receive antenna case and extended to the
MIMO case in~\cite{SSJB05,KTA06}.  These MSE and SINR dualities are
equally applicable to sum rate maximization.

Linear precoding approaches to sum rate maximization have been
proposed for both single-antenna receivers~\cite{SVH06,SJHU06} and for
multiple antenna receivers~\cite{CTJL07,SSB07,TA06}.  In~\cite{SVH06},
the authors suggest an iterative method for direct optimization of the
sum rate, while~\cite{SJHU06} and~\cite{CTJL07} exploit the SINR
uplink-downlink duality of~\cite{SB04,SSJB05,KTA06}.  In~\cite{SSB07}
and~\cite{TA06}, two similar algorithms were independently proposed to
minimize the product of the mean squared errors (PMSE) in the
multiuser MIMO downlink; these papers showed that the PMSE
minimization problem is equivalent to the direct sum rate maximization
proposed in~\cite{SJHU06,SVH06,CTJL07}.  The work of~\cite{TA06} was
motivated by the equivalence relationship developed between the single
user minimum MSE (MMSE) and mutual information in~\cite{GSV05}.  Each of the approaches
in~\cite{SVH06,SJHU06,CTJL07,SSB07,TA06} yields a suboptimal solution,
as the resulting solutions converge only to a local optimum, if at
all.

Given this prior work in linear precoding, an important motivation for
this paper is to determine the performance upper bound achievable
under linear precoding and to evaluate how closely PMSE minimization
comes to approaching this upper bound.  In the single-user multicarrier case,
minimizing the PMSE is equivalent to minimizing the determinant of the
MSE matrix and thus is also equivalent to maximizing the mutual
information~\cite{PCL03}.  This equivalence does not apply to the multiuser scenario. 
In this paper, we investigate the relationship
between the MSE-matrix determinants, the mutual information, and the
maximum achievable sum rate under linear precoding in the multiuser
MIMO downlink, resulting in an optimization problem based on
minimizing the product of the determinants of all users' MSE matrices
(PDetMSE).  Furthermore, we underline the differences between the
joint (multi-stream) optimization that arises from the PDetMSE
approach and the scalar (per-stream) PMSE-based solution.
While chronologically, the PMSE approach was developed before the PDetMSE
formulation, we present PMSE in this paper as a lower complexity
approximation of the PDetMSE formulation.

The main contributions of this paper are:
\begin{itemize}
\item Deriving the maximum achievable information rates for both joint
  and scalar processing under linear precoding and formulating the
  joint (PDetMSE) and scalar processing (PMSE) based sum rate
  maximization problems using MSE expressions.
\item Proposing solutions to these optimization problems based on
  uplink-downlink duality, and addressing several issues regarding
  algorithm implementation.
\item Analyzing the performance of our proposed schemes in comparison
  to the DPC sum capacity and to orthogonalization based approaches.
  We demonstrate that a performance improvement is made in narrowing
  the gap to capacity at practical values of transmit SNR, and show
  that the PDetMSE approach provides the best performance of all
  proposed schemes.
\end{itemize}

The remainder of this paper is organized as
follows. Section~\ref{section:model} describes the system model used
and states the assumptions made.  Section~\ref{section:sumcaplp}
derives the performance upper bound for the achievable sum rate under
linear precoding, and develops the use of the product of MSE matrix
determinants as the optimization criterion for joint processing.
Section~\ref{section:subopt} investigates a suboptimal framework based
on the product of mean squared errors and proposes a computationally feasible scheme for
implementation.  Results of simulations testing the
effectiveness of the proposed approaches are presented in
Section~\ref{section:sims}. Finally, we draw our conclusions in
Section~\ref{section:conc}.

\emph{Notation}: Lower case italics, e.g.,~$x$, represent scalars while
lower case boldface type is used for vectors (e.g., $\mathbf{x}$).
Upper case italics, e.g.,~$N$, are used for constants and upper case
boldface represents matrices, e.g., $\mathbf{X}$.  Entries in vectors
and matrices are denoted as $\left[\mathbf{x}\right]_i$ and
$\left[\mathbf{X}\right]_{i,j}$ respectively. The superscripts $^T$ and
$^H$ denote the transpose and Hermitian operators. $\mathbb{E}[\cdot]$
represents the statistical expectation operator while $\mathbf{I}_N$ is
the $N\times N$ identity matrix. $\mathrm{tr}\left[\cdot \right]$ and
$\mathrm{det}\left(\cdot \right)$ are the trace and determinant
operators.  $\left\|\mathbf{x}\right\|_1$ and $\left\|\mathbf{x}
\right\|_2$ denote the 1-norm (sum of entries) and Euclidean norm.
$\mathrm{diag}(\mathbf{x})$ represents the diagonal matrix formed using
the entries in vector $\mathbf{x}$, and
$\mathrm{diag}\left[\mathbf{X}_1,\ldots,\mathbf{X}_k \right]$ is the
block diagonal concatenation of matrices
$\mathbf{X}_1,\ldots,\mathbf{X}_k$. $\mathbf{A}\succ \mathbf{0}$ and
$\mathbf{B} \succeq \mathbf{0}$ indicate that $\mathbf{A}$ and
$\mathbf{B}$ are positive definite and positive semidefinite matrices,
respectively.  $\hat{e}_{\max}(\mathbf{A},\mathbf{B})$ is the unit Euclidean norm
eigenvector $\mathbf{x}$ corresponding to the largest eigenvalue
$\lambda$ in the generalized eigenproblem
$\mathbf{Ax}=\lambda\mathbf{Bx}$. Finally, $\mathcal{CN}(m,\sigma^2)$
denotes the complex Gaussian probability distribution with mean $m$ and
variance $\sigma^2$.

\section{System Model with Linear Precoding\label{section:model}}
\begin{figure}[!t]
\centering
\includegraphics[width=\plotwidth]{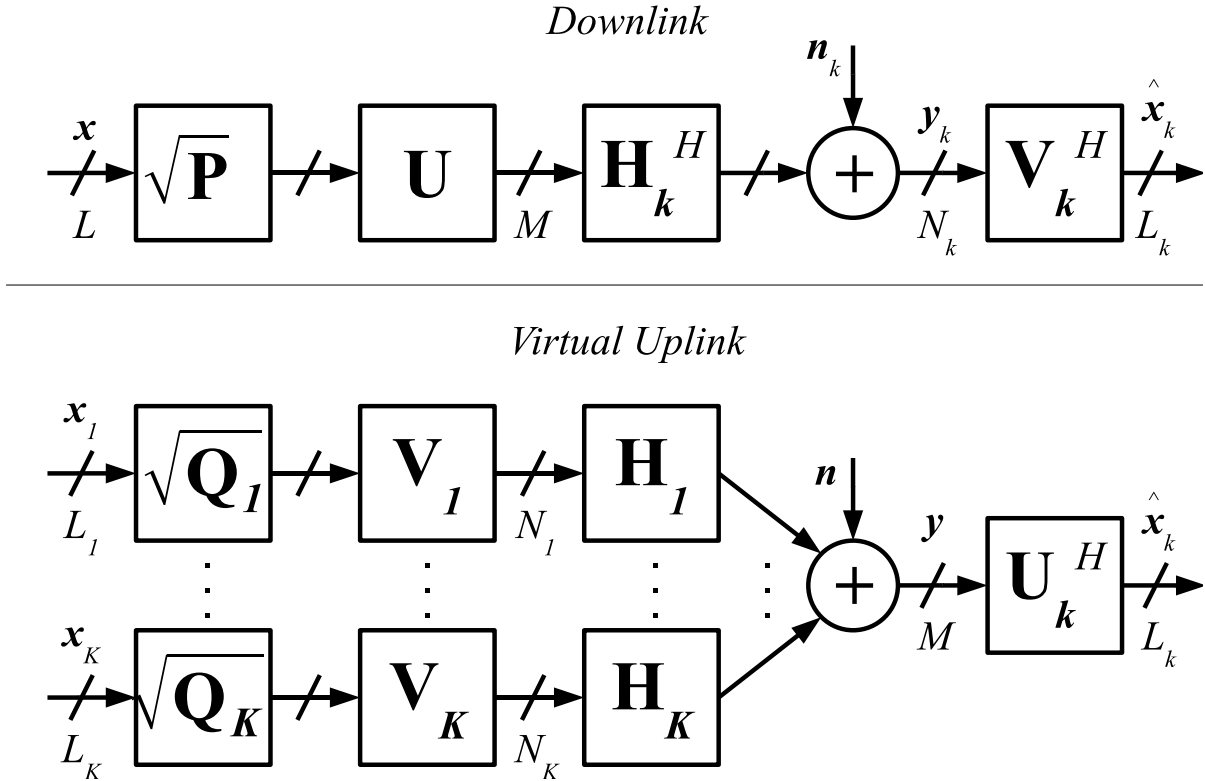}
\caption{Processing for user $k$ in downlink and virtual uplink.}\label{fig:scheme}
\end{figure}

The system under consideration, illustrated in Fig.~\ref{fig:scheme}, comprises
a base station with $M$ antennas transmitting to $K$ decentralized users over
flat wireless channels.
User $k$ is equipped with $N_{k}$ antennas and receives
$L_k$ data streams from the base station. Thus, we have $M$ transmit antennas
transmitting a total of $L=\sum_{k=1}^{K}L_{k}$ symbols to $K$ users, who,
together, have a total of $N=\sum_{k=1}^{K}N_{k}$ receive antennas. The data
symbols for user $k$ are collected in the data vector
$\mathbf{x}_{k}=\left[x_{k1}, x_{k2}, \ldots, x_{kL_{k}}\right]^T$ and the
overall data vector is $\mathbf{x} = \left[\mathbf{x}_1^T, \mathbf{x}_2^T,
\ldots, \mathbf{x}_K^T \right]^T$.   We assume that the modulated data symbols $\mathbf{x}$ are independent with unit average energy ($\mathbb{E}\left[\mathbf{x}\mathbf{x}^{H}\right]=\mathbf{I}_{L}$). User $k$'s data streams are processed by the $M\times L_k$ transmit filter $\mathbf{U}_{k} = \left[ \mathbf{u}_{k1}, \ldots, \mathbf{u}_{kL_k}\right]$ before being transmitted over the $M$
antennas; $\mathbf{u}_{kj}$ is the precoder for stream $j$ of user $k$, and has
unit power $\|\mathbf{u}_{kj}\|_2 = 1$.  Together, these individual precoders
form the $M\times L$ global transmitter precoder matrix $\mathbf{U} =
\left[\mathbf{U}_{1}, \mathbf{U}_{2}, \ldots,
  \mathbf{U}_{K}\right]$. Let
$p_{kj}$ be the power allocated to stream $j$ of user  $k$ and the
downlink transmit power vector for user $k$ be
$\mathbf{p}_{k}=\left[p_{k1}, p_{k2}, \ldots, p_{kL_{k}} \right]^T$, with
$\mathbf{p}=\left[ \mathbf{p}_1^T, \ldots, \mathbf{p}_K^T \right]^T$.
Define $\mathbf{P}_{k}=\mathrm{diag}\{\mathbf{p}_{k}\}$ and
$\mathbf{P}=\mathrm{diag}\{\mathbf{p}\}$. The channel between the
transmitter and user $k$ is represented by the $N_{k}\times M$ matrix
$\mathbf{H}_{k}^{H}$.  The overall
$N\times M$ channel matrix is $ \mathbf{H}^H$, with
$\mathbf{H}=\left[\mathbf{H}_{1},\,\mathbf{H}_{2},\,\dots,\mathbf{H}_{K}
  \right]$. The transmitter is assumed to know the channel perfectly.

Based on this model, user $k$ receives a length-$N_{k}$ vector
\begin{equation*}
\mathbf{y}_{k}=\mathbf{H}_{k}^{H}\mathbf{U}\sqrt{\mathbf{P}}\mathbf{x} +
\mathbf{n}_{k},
\end{equation*}
where $\mathbf{n}_k$ consists of the additive white Gaussian noise (AWGN)
at the user's receive antennas with i.i.d.\ entries
$\left[\mathbf{n}_k\right]_i \sim \mathcal{CN}(0,\sigma^2$); that is,
$\mathbb{E}\left[\mathbf{n}_{k}\mathbf{n}_{k}^{H}\right]=\sigma^2
\mathbf{I}_{N_{k}}$.  To estimate its $L_{k}$ symbols $\mathbf{x}_{k}$, user $k$ processes $\mathbf{y}_{k}$ with its
$L_{k}\times N_{k}$ decoder matrix $\mathbf{V}_{k}^{H}$ resulting in
\begin{equation*}
\hat{\mathbf{x}}_{k}^{DL}=\mathbf{V}_{k}^{H}\mathbf{H}_{k}^{H}\mathbf{U}
\sqrt{\mathbf{P}}\mathbf{x}+\mathbf{V}_{k}^{H}\mathbf{n}_{k},
\end{equation*}
where the superscript $^{DL}$ indicates the downlink.  The global receive filter $\mathbf{V}^{H}$ is a block diagonal
matrix of dimension $L\times N$, $\mathbf{V} =
\mathrm{diag}\left[\mathbf{V}_{1}, \, \mathbf{V}_{2}, \cdots,
\mathbf{V}_{K}\right]$, where each $\mathbf{V}_{k} =
\left[ \mathbf{v}_{k1}, \ldots,
\mathbf{v}_{kL_k}\right]$.  The MSE matrix for user $k$ in the downlink under these general precoder and decoder matrices can be written as
\begin{equation}\label{eqn:msedlmodel}
\begin{split}
\mathbf{E}^{DL}_k & = \mathbb{E}\left[\left(\hat{\mathbf{x}}_k -
\mathbf{x}_k\right)\left(\hat{\mathbf{x}}_k - \mathbf{x}_k\right)^H\right] \\
& = \mathbf{V}_{k}^{H}\mathbf{H}_{k}^{H}\mathbf{U}\mathbf{P}\mathbf{U}^{H}\mathbf{H}_{k}\mathbf{V}_{k} + \sigma^2\mathbf{V}_k^{H}\mathbf{V}_k \\
& \quad -\mathbf{V}_{k}^{H}\mathbf{H}_{k}^{H}\mathbf{U}_k\sqrt{\mathbf{P}_k}  -\sqrt{\mathbf{P}_k}\mathbf{U}_k^{H}\mathbf{H}_{k}\mathbf{V}_{k} + \mathbf{I}_{L_k}.
\end{split}
\end{equation}

We will make use of the dual virtual uplink, also illustrated in Fig.~\ref{fig:scheme}, with the same channels between users and base station.
In the uplink, user $k$ transmits $L_k$ data streams.  Let the uplink transmit power vector for user $k$ be $\mathbf{q}_{k}=[q_{k1}, q_{k2}, \ldots, q_{kL_{k}}]^T$, with
$\mathbf{q}=[\mathbf{q}_{1}^T,\ldots,\mathbf{q}_{K}^T]^T$, and define
$\mathbf{Q}_{k}=\mathrm{diag}\{\mathbf{q}_{k}\}$ and
$\mathbf{Q}=\mathrm{diag}\{\mathbf{q}\}$.  The transmit and receive filters for
user $k$ become $\mathbf{V}_{k}$ and $\mathbf{U}_{k}^{H}$ respectively. As in
the downlink, the precoder for the virtual uplink contains columns with unit
norm; that is, $\|\mathbf{v}_{kj}\|_2 = 1$. The received vector at the base
station and the estimated symbol vector for user $k$ are
\begin{align*}
\mathbf{y} & = \sum_{i=1}^{K}\mathbf{H}_{i}\mathbf{V}_{i}\sqrt{\mathbf{Q}_{i}}
\mathbf{x}_{i}+\mathbf{n},\\
\hat{\mathbf{x}}_{k}^{UL}& = \sum_{i=1}^{K}\mathbf{U}_{k}^{H}\mathbf{H}_{i}
\mathbf{V}_{i}\sqrt{\mathbf{Q}_{i}}\mathbf{x}_{i}+
                        \mathbf{U}_{k}^{H}\mathbf{n}.
\end{align*}
The noise term, $\mathbf{n}$, is again AWGN with
$\mathbb{E}\left[\mathbf{n}\mathbf{n}^{H}\right]=\sigma^{2}\mathbf{I}_{M}$.

We define a useful virtual uplink receive covariance matrix as
\begin{align*}
\mathbf{J} = \mathbb{E}\left[\mathbf{y}\mathbf{y}^{H}\right]
& = \sum_{k=1}^K \mathbf{H}_k\mathbf{V}_k\mathbf{Q}_k\mathbf{V}_k^H\mathbf{H}_k^H
                                                   + \sigma^2\mathbf{I}_M\\
                                                   &  = \mathbf{HVQV}^H\mathbf{H}^H + \sigma^2 \mathbf{I}_M.
\end{align*}
The global MSE matrix for all users in the virtual uplink can then be expressed as
\begin{equation}\label{eqn:mseulmodel}
\begin{split}
\mathbf{E}^{UL} & = \mathbb{E}\left[\left(\hat{\mathbf{x}} -
\mathbf{x}\right)\left(\hat{\mathbf{x}} - \mathbf{x}\right)^H\right] \\
& = \mathbf{U}^H\mathbf{J}\mathbf{U} -\mathbf{U}^{H}\mathbf{HV}\sqrt{\mathbf{Q}}-\sqrt{\mathbf{Q}}\mathbf{V}^{H}\mathbf{H}^{H}\mathbf{U} + \mathbf{I}_{L}.
\end{split}
\end{equation}

\section{Linear Precoding and Sum Rate Maximization \label{section:sumcaplp}}
In this section, we formulate the sum rate maximization problem under
linear precoding in the broadcast channel.  We begin by introducing
the information theoretic DPC upper bound, and then derive the
performance upper bound achievable under linear precoding.  We then
derive an equivalent formulation in terms of MSE expressions, and
propose the PDetMSE based scheme for achieving this optimal sum rate
performance under linear precoding.

\subsection{Sum Capacity and Dirty Paper Coding\label{sub:bcsumcap}}
Information theoretic approaches characterize the sum capacity of the multiuser MIMO downlink by solving the sum capacity of the equivalent uplink multiple access channel (MAC) and applying a duality result~\cite{VT03, VJG03}.  The BC sum capacity can thus be expressed as
\begin{align*}
R_{\mathrm{sum}} & = \max_{\mathbf{\Sigma}_k} \log \det
\left( \mathbf{I} + \frac{1}{\sigma^2}
      \sum_{k=1}^K \mathbf{H}_k \mathbf{\Sigma}_k \mathbf{H}_k^H \right) \\
& \mathrm{s.t.} \quad \mathbf{\Sigma}_k \succeq \mathbf{0}, \quad k = 1, \ldots, K\\
& \quad \quad \sum_{k=1}^K \mathrm{tr}\left[ \mathbf{\Sigma}_k\right] \le
P_\mathrm{max},
\end{align*}
where $\mathbf{\Sigma}_k$ is the uplink transmit covariance matrix for
mobile user $k$, and $P_{\max}$ is the maximum sum power over all
users.  Note that this optimization problem is concave in
$\mathbf{\Sigma}_k$, and is hence relatively easy to solve.  This
result does not translate to linear precoding.

\subsection{Achievable Sum Rate under Linear Precoding\label{sub:linearprecoding}}
The achievable rate for a single user MIMO channel is $\log\left(\det{\left(\mathbf{K}_x + \mathbf{K}_z\right)} / \det{\left(\mathbf{K}_z\right)}\right)$ (where $\mathbf{K}_x$ is the received signal covariance and $\mathbf{K}_z$ is the noise covariance)~\cite{CT06}.  Under single-user decoding, multi-user interference is treated as noise, and user $k$ can achieve rate $R_k$ in the downlink using transmit covariance $\mathbf{\Sigma}_k$:
\begin{equation*}
R_k = \log \frac{\det{\left( \sum_{j=1}^K \mathbf{H}_k^H \mathbf{\Sigma}_j \mathbf{H}_k + \sigma^2 \mathbf{I} \right)}}
{\det{\left( \sum_{j \ne k} \mathbf{H}_k^H \mathbf{\Sigma}_j \mathbf{H}_k + \sigma^2 \mathbf{I} \right)}}.
\end{equation*}

Under the system model described in Section~\ref{section:model}, user $k$ transmits with covariance matrix $\mathbf{\Sigma}_k = \mathbf{U}_k\mathbf{P}_k\mathbf{U}_k^H$.  The achievable rate for user $k$ under linear precoding is therefore
\begin{equation}\label{eqn:linearAchievable}
\begin{split}
R^{\mathbf{LP}}_k & = \log \frac{\det{\left( \sum_{j=1}^K \mathbf{H}_k^H \mathbf{U}_j \mathbf{P}_j \mathbf{U}_j^H \mathbf{H}_k + \sigma^2 \mathbf{I} \right)}}
{\det{\left( \sum_{j \ne k} \mathbf{H}_k^H \mathbf{U}_j \mathbf{P}_j \mathbf{U}_j^H \mathbf{H}_k + \sigma^2 \mathbf{I} \right)}}\\
& = \log \frac{\det{\mathbf{J}_k}}{\det{\mathbf{R}_{N+I,k}}},
\end{split}
\end{equation}
where $\mathbf{J}_k = \mathbf{H}_k^H\mathbf{UPU}^H\mathbf{H}_k +
\sigma^2\mathbf{I}$ and $\mathbf{R}_{N+I,k} = \mathbf{J}_k -
\mathbf{H}_k^H\mathbf{U}_k\mathbf{P}_k\mathbf{U}_k^H\mathbf{H}_k$ are
the received signal covariance matrix and the noise-plus-interference covariance matrix at user $k$, respectively.

The rate maximization problem with a sum power constraint under linear
precoding can then be formulated as
\begin{eqnarray}
\nonumber (\mathbf{U},\mathbf{P}) & = &  \arg\max_{\mathbf{U},\mathbf{P}}
  \sum_{k=1}^K \log
  \frac{\det{\mathbf{J}_k}}{\det{\mathbf{R}_{N+I,k}}}\\
\nonumber & \mathrm{s.t.} & \|\mathbf{u}_{kj}\|_2 = 1, \quad k = 1, \ldots, K,\quad j=1,\ldots,L_k\\
\nonumber & & p_{kj} \ge 0, \hspace{0.36in} k=1,\ldots,K,\quad j = 1, \ldots, L_k\\
& & \|\mathbf{p}\|_1 = \sum_{k=1}^K\sum_{j=1}^{L_k} p_{kj} \le
P_{\mathrm{max}}.\label{eqn:rateMax}
\end{eqnarray}

\subsection{MSE Formulation: Product of MSE Matrix Determinants\label{sub:pdetmse}}
In this section, we show that an MSE-based formulation using joint
processing of all streams (rather than treating each user's own data streams as interference) leads to
an equivalent optimal formulation of the rate maximization problem
under linear processing. We develop this relationship by using the MSE matrix determinants.

First, consider the linear MMSE decoder for user $k$, $\mathbf{V}_k$,
\begin{equation}\label{eqn:mmserx}
\begin{split}
\mathbf{V}_k & = \left(\mathbf{H}_k^H\mathbf{UPU}^H\mathbf{H}_k +
\sigma^2\mathbf{I}\right)^{-1}\mathbf{H}_k^H\mathbf{U}_k\sqrt{\mathbf{P}_k}\\
& = \mathbf{J}_k^{-1}\mathbf{H}_k^H\mathbf{U}_k\sqrt{\mathbf{P}_k}.
\end{split}
\end{equation}
When using this matrix as the receiver in (\ref{eqn:msedlmodel}), the
downlink MSE matrix for user $k$ in can be simplified as
\begin{equation}\label{eqn:msedl1}
\mathbf{E}^{DL}_k = \mathbf{I}_{L_k} - \sqrt{\mathbf{P}_k}\mathbf{U}_k^H\mathbf{H}_k
\mathbf{J}_k^{-1}\mathbf{H}_k^H\mathbf{U}_k\sqrt{\mathbf{P}_k}.
\end{equation}

Consider the following optimization problem which minimizes the
product of the determinants of the downlink MSE matrices under a sum
power constraint:
\begin{eqnarray}
\nonumber (\mathbf{U},\mathbf{P}) & = &
  \arg\min_{\mathbf{U},\mathbf{P}}\prod_{k=1}^K \det \mathbf{E}^{DL}_k\\
\nonumber & \mathrm{s.t.} & \|\mathbf{u}_{kj}\|_2 = 1, \quad k = 1, \ldots, K, \quad j = 1, \ldots, L_k\\
\nonumber & & p_{kj} \ge 0, \hspace{0.36in} k = 1, \ldots, K, \quad j = 1, \ldots, L_k\\
& & \|\mathbf{p}\|_1 = \sum_{k=1}^K\sum_{j=1}^{L_k} p_{kj} \le
P_{\mathrm{max}}.\label{eqn:logdetmin}
\end{eqnarray}

\begin{theorem}
Under linear MMSE decoding at the base station, the sum rate maximization problem in (\ref{eqn:rateMax}) and the PDetMSE minimization problem in (\ref{eqn:logdetmin}) are equivalent.
\end{theorem}

\begin{IEEEproof}
The determinant of the downlink MSE matrix can be written as
\begin{align}
\det \mathbf{E}^{DL}_k & = \det\left(\mathbf{I}_{L_k} - \mathbf{H}_k^H\mathbf{U}_k\mathbf{P}_k\mathbf{U}_k^H\mathbf{H}_k
\mathbf{J}_k^{-1}\right)\label{eqn:detident1}\\
& = \det \left[\left(\mathbf{J}_k -
  \mathbf{H}_k^H\mathbf{U}_k\mathbf{P}_k\mathbf{U}_k^H\mathbf{H}_k
  \right)\mathbf{J}_k^{-1}\right]\nonumber\\
& = \det \left[\mathbf{R}_{N+I,k}\mathbf{J}_k^{-1}\right]\nonumber\\
& = \frac{\det\mathbf{R}_{N+I,k}}{\det \mathbf{J}_k},\nonumber
\end{align}
where (\ref{eqn:detident1}) follows from (\ref{eqn:msedl1}) since
$\det(\mathbf{I}+\mathbf{AB}) = \det(\mathbf{I}+\mathbf{BA})$ when
$\mathbf{A}$ and $\mathbf{B}$ have appropriate dimensions.  We then
see the relationship to (\ref{eqn:linearAchievable}),
\begin{equation*}
\begin{split}
\log \det \mathbf{E}^{DL}_k & = - \log\frac{\det
  \mathbf{J}_k}{\det\mathbf{R}_{N+I,k}}\\
& = -R^{\mathbf{LP}}_k.
\end{split}
\end{equation*}
With this result, we can see that under MMSE reception using
$\mathbf{V}_k$ as defined in~(\ref{eqn:mmserx}), minimizing
the determinant of the MSE matrix $\mathbf{E}^{DL}_k$ is equivalent to
maximizing the achievable rate for user $k$.  It follows that
minimizing the product of MSE matrix determinants over all users is
equivalent to sum rate maximization,
\begin{align}
\min \prod_{k=1}^K \det \mathbf{E}^{DL}_k & \equiv \min \sum_{k=1}^K
  \log \det \mathbf{E}^{DL}_k\label{eqn:mserateequiv}\\
& \equiv \max \sum_{k=1}^K R^{\mathbf{LP}}_k.\nonumber
\end{align}
where (\ref{eqn:mserateequiv}) holds since since
$\log(\cdot)$ is a monotonically increasing function of its argument.
\end{IEEEproof}
Note that this new result represents an upper bound on the sum rate on all
linear precoding schemes in the broadcast channel.

The covariance matrices $\mathbf{J}_k$ and $\mathbf{R}_{N+I,k}$ in the MSE
matrix $\mathbf{E}_k$ are each functions of all precoder and power allocation
matrices.  Thus, the sum rates $R_k$ for each user $k$ (and the sum rate for
all users) are coupled across users.  As such, finding $\mathbf{U}$ and
$\mathbf{P}$ jointly or finding only the power allocation $\mathbf{P}$ for a
fixed $\mathbf{U}$ are both non-convex problems and are just as difficult to
solve as the rate maximization problem.

In the sum capacity and SMSE problems, the problem of non-convexity is
addressed by solving a convex virtual uplink formulation and applying
a duality-based transformation.  Unfortunately, the sum rate
expression under linear precoding in the virtual uplink is nearly
identical to that derived above for the downlink, and does not admit a
cancellation or grouping of terms to decouple the problem across
users.  

Direct solution of the non-convex downlink problem for minimizing the product
of MSE matrix determinants requires finding a complex $M \times L$ precoder
matrix.  We consider the application of sequential quadratic
programming (SQP)~\cite{BT95} to solve this problem.  SQP solves
successive approximations of a constrained optimization problem and is
guaranteed to converge to the optimum value
for convex problems; however, in the case of this non-convex optimization
problem, SQP can only guarantee convergence to a local minimum.

This computationally intensive approach is the only available option
in the absence of a convex virtual uplink formulation.  Moreover, the 
numerical techniques used for solving nonlinear problems do not guarantee convergence to the global minimum.  This is clearly not a desirable method for finding a practical precoder, especially when one of
our major motivations for using linear precoding is reducing transmitter complexity.  We do not suggest that this method be practically implemented; rather, we use it to illustrate the difference in performance between the solutions to the optimal PDetMSE formulation and the more practical PMSE
algorithm that we propose in the following section.

\section{Scalar Processing and the Product of Mean Squared Errors\label{section:subopt}}
Given the complexity of the PDetMSE solution, we consider PMSE
minimization as a suboptimal (but likely feasible) approximation
to rate maximization in the multiuser MIMO downlink. In~\cite{PCL03}, the single-user rate maximization problem using linear precoding is solved by minimizing the determinant of the MSE matrix.  This solution 
is equivalent to minimizing the product of individual stream MSEs because the problem is scalarized by diagonalization of both the channel and MSE 
matrices.  It was recently demonstrated in~\cite{HJ08} that the MSE matrices can also be diagonalized in the multiuser case by applying unitary transformations 
to the precoder and decoders; however, in the absence of orthogonalizing precoders (e.g., BD or ZF), minimization of the PMSE yields a different solution from 
minimizing the PDetMSE.

The PMSE approach, based on scalar processing of the individual stream MSEs for each user,
follows from the treatment of the optimization problems
in~\cite{SSJB05,KTA06}, where non-convex problems in the downlink are
transformed to convex problems in the dual uplink.  With this
motivation in mind, we consider formulating the scalar optimization
problem directly in the virtual uplink, and transforming the resulting
solution to the downlink using the uplink-downlink MSE duality
in~\cite{SSJB05,KTA06}.

\subsection{Achievable Sum Rate using Scalar Processing}
In the scalarized version of the rate maximization problem, the user's
own data streams $(l \ne j)$ are considered as self-interference in
addition to the multiuser interference.  The achievable rate for user
$k$'s substream $j$ can thus be expressed as
\begin{equation*}
R^{\mathbf{LP}}_{k,j} = \log \left(1 + \gamma_{kj}^{UL}\right),
\end{equation*}
where
\begin{equation}\label{eqn:SINRUL}
\gamma_{kj}^{UL} = \frac{\mathbf{u}_{kj}^H\mathbf{H}_k\mathbf{v}_{kj}
q_{kj}\mathbf{v}_{kj}^H\mathbf{H}_k^H\mathbf{u}_{kj}}
{\mathbf{u}_{kj}^H\mathbf{J}_{kj}\mathbf{u}_{kj}}
\end{equation}
is the SINR and $\mathbf{J}_{kj} = \mathbf{J} -
\mathbf{H}_k\mathbf{v}_{kj}q_{kj}\mathbf{v}_{kj}^H\mathbf{H}_k^H$ is
the virtual uplink interference-plus-noise covariance matrix for stream $j$ of user $k$.

The scalar rate maximization problem with a sum power constraint under linear
precoding can thus be written as
\begin{eqnarray}
\nonumber (\mathbf{V},\mathbf{Q}) & = &  \arg\max_{\mathbf{V},\mathbf{Q}}
  \sum_{k=1}^K \sum_{j=1}^{L_k} \log \left( 1+\gamma_{kj}^{UL} \right)\\
\nonumber & \mathrm{s.t.} & \|\mathbf{v}_{kj}\|_2 = 1, \quad k = 1, \ldots, K, \quad j = 1,\ldots,L_k\\
\nonumber & & q_{kj} \ge 0, \hspace{0.36in} k=1,\ldots,K,\quad j = 1, \ldots, L_k\\
& & \|\mathbf{q}\|_1 = \sum_{k=1}^K\sum_{j=1}^{L_k} q_{kj} \le
P_{\mathrm{max}}\label{eqn:ratemaxscalar}.
\end{eqnarray}

\subsection{MSE Formulation: Product of Mean Squared Errors}
With this scalar processing rate maximization problem in mind, we
consider the MSE-equivalent formulation.  We begin by finding the
optimum linear receiver, and can see from (\ref{eqn:SINRUL}) that
$\mathbf{u}_{kj}$ does not depend on any other columns of $\mathbf{U}$.
Furthermore, it is the solution to the generalized eigenproblem
\begin{equation*}
\mathbf{u}_{kj}^{\mathrm{opt}} = \hat{e}_{\max}\left(\mathbf{H}_k\mathbf{v}_{kj}q_{kj}
\mathbf{v}_{kj}^H\mathbf{H}_k^H, \mathbf{J}_{kj} \right).
\end{equation*}
Within a normalizing factor, this
solution is equivalent to the MMSE receiver,
\begin{equation}\label{eqn:u_mmse}
\mathbf{u}_{kj}^{\mathrm{opt}} = \mathbf{J}^{-1}\mathbf{H}_k\mathbf{v}_{kj}\sqrt{q_{kj}}.
\end{equation}
When the MMSE receiver in (\ref{eqn:u_mmse}) is used, the virtual uplink MSE matrix (\ref{eqn:mseulmodel}) reduces to
\begin{equation*}
\mathbf{E} ^{UL} = \mathbf{I}_{L} - \sqrt{\mathbf{Q}}\mathbf{V}^H\mathbf{H}^H
\mathbf{J}^{-1}\mathbf{HV}\sqrt{\mathbf{Q}}.
\end{equation*}
Thus, the mean squared error for user $k$'s $j^{\mathrm{th}}$ stream is entry $j$ in block $k$ of $\mathbf{E}^{UL}$,
\begin{equation*}
\epsilon_{kj}^{UL} = 1 - q_{kj} \mathbf{v}_{kj}^H\mathbf{H}_k^H
                             \mathbf{J}^{-1}\mathbf{H}_k\mathbf{v}_{kj}.
\end{equation*}

Now consider another optimization problem, minimizing the product of mean
squared errors (PMSE) under a sum power constraint,
\begin{eqnarray}\label{eqn:minpmse}
\nonumber\left(\mathbf{V}, \mathbf{Q}\right) & = &
\arg\min_{\mathbf{V}, \mathbf{Q}} \prod_{k=1}^K\prod_{j=1}^{L_k}\epsilon_{kj}^{UL}\\
\nonumber & \mathrm{s.t.} & \|\mathbf{v}_{kj}\|_2 = 1, \quad k = 1, \ldots, K, \quad j = 1, \ldots, L_k\\
\nonumber & & q_{kj} \ge 0, \hspace{0.36in} k = 1, \ldots, K, \quad j = 1, \ldots, L_k\\
& & \|\mathbf{q}\|_1 = \sum_{k=1}^K\sum_{j=1}^{L_k} q_{kj} \le
P_{\mathrm{max}}.
\end{eqnarray}

\begin{theorem}
Under linear MMSE decoding at the base station, the optimization problems
defined by (\ref{eqn:ratemaxscalar}) and (\ref{eqn:minpmse}) are equivalent.
\end{theorem}

\begin{IEEEproof}
Using (\ref{eqn:SINRUL}), we can rewrite $1 + \gamma_{kj}^{UL}$ as
\begin{equation*}
 1 + \gamma_{kj}^{UL} = \frac{\mathbf{u}_{kj}^H\mathbf{J}\mathbf{u}_{kj}}
{\mathbf{u}_{kj}^H\mathbf{J}\mathbf{u}_{kj} -
\mathbf{u}_{kj}^H\mathbf{H}_k\mathbf{v}_{kj}q_{kj}
\mathbf{v}_{kj}^H\mathbf{H}_k^H\mathbf{u}_{kj}}.
\end{equation*}
It follows that by using the MMSE receiver from (\ref{eqn:u_mmse}),
\begin{equation}\label{eqn:invSINR}
\begin{split}
\frac{1}{1 + \gamma_{kj}^{UL}} & = 1 - \frac{\mathbf{u}_{kj}^H\mathbf{H}_k
   \mathbf{v}_{kj}q_{kj}\mathbf{v}_{kj}^H\mathbf{H}_k^H
    \mathbf{u}_{kj}}{\mathbf{u}_{kj}^H\mathbf{J}
    \mathbf{u}_{kj}}\\
& = 1 - \frac{ \left(q_{kj}\mathbf{v}_{kj}^H\mathbf{H}_k^H
 \mathbf{J}^{-1}\mathbf{H}_k\mathbf{v}_{kj}\right)^2}{q_{kj}\mathbf{v}_{kj}^H
     \mathbf{H}_k^H\mathbf{J}^{-1}\mathbf{H}_k\mathbf{v}_{kj}}\\
& = 1 - q_{kj}\mathbf{v}_{kj}^H\mathbf{H}_k^H\mathbf{J}^{-1}
           \mathbf{H}_k\mathbf{v}_{kj} = \epsilon_{kj}^{UL}.
\end{split}
\end{equation}
This relationship is similar to one shown for MMSE detection in
CDMA systems~\cite{MH94}.  By applying (\ref{eqn:invSINR}) to
(\ref{eqn:ratemaxscalar}), we see that
\begin{equation*}
\sum_{k=1}^K\sum_{j=1}^{L_k} \log \left( 1 + \gamma_{kj}^{UL} \right )
= -\log \left( \prod_{k=1}^K\prod_{j=1}^{L_k} \epsilon_{kj}^{UL} \right ).
\end{equation*}

Since the constraints on $\mathbf{v}_{kj}$ and $q_{kj}$ are identical in
(\ref{eqn:ratemaxscalar}) and (\ref{eqn:minpmse}), the problem of maximizing sum
rate in (\ref{eqn:ratemaxscalar}) is therefore equivalent to minimizing the PMSE in
(\ref{eqn:minpmse}).
\end{IEEEproof}

Note that this result has been independently derived in~\cite{SSB07,TA06}.

\subsection{Algorithm: PMSE Minimization\label{section:pmse_alg}}
We now present an algorithm that minimizes the product of mean squared errors.  The
algorithm draws upon previous work based on uplink-downlink MSE duality~\cite{SSJB05,KTA06}, which states that all achievable MSEs in the \emph{uplink} for a given $\mathbf{U}$, $\mathbf{V}$, and
$\mathbf{q}$ (with sum power constraint $\|\mathbf{q}\|_1 \le P_{\max}$),
can also be achieved by a power allocation $\mathbf{p}$ in the
\emph{downlink} (where $\|\mathbf{p}\|_1 \le P_{\max}$).
It operates by iteratively obtaining the downlink precoder matrix
$\mathbf{U}$ and power allocations $\mathbf{p}$ and the virtual uplink
precoder matrix $\mathbf{V}$ and power allocations $\mathbf{q}$. Each
step minimizes the objective function by modifying one of these four
variables while leaving the remaining three fixed.

\subsubsection{Downlink Precoder}
For a fixed set of virtual uplink precoders $\mathbf{V}_k$ and power
allocation $\mathbf{q}$, the optimum virtual uplink decoder $\mathbf{U}$
is defined by (\ref{eqn:u_mmse}).  Each $\epsilon_{kj}$ is minimized
individually by this MMSE receiver, thereby also minimizing the product
of MSEs.  This $\mathbf{U}$ is normalized and used as the downlink precoder.

\subsubsection{Downlink Power Allocation}
The downlink power allocation $\mathbf{p}$ is given by~\cite{KTA06}:
\begin{equation*}
\mathbf{p}=\sigma^{2}(\mathbf{D}^{-1}-\mathbf{\Psi})^{-1}\mathbf{1},
\end{equation*}
where $\mathbf{\Psi}$ is the $L\times L$ cross coupling matrix defined as
\begin{equation*}
[\mathbf{\Psi}]_{ij}=\left\{ \begin{array}{ll}
|\mathbf{\tilde{h}}_{i}^{H}\mathbf{u}_{j}|^{2}=|\mathbf{u}_{j}^{H}
\mathbf{\tilde{h}}_{i}|^{2} & \textrm{${i}\neq{j}$}\\
0 & \textrm{$i=j$}\end{array} \right.,
\end{equation*}
\begin{equation*}
\mathbf{D}=\mathrm{diag}\left\{\frac{\gamma_{11}^{UL}}{|\mathbf{v}_{11}^{H}\mathbf{H}_{1}^{H}
\mathbf{u}_{11}|^2},\dots,
\frac{\gamma_{KL_{K}}^{UL}}{|\mathbf{v}_{KL_{K}}^{H}\mathbf{H}_{K}^{H}
\mathbf{u}_{KL_{K}}|^2}\right\},
\end{equation*}
where $\mathbf{\tilde{H}}=\mathbf{H}\mathbf{V}=[\mathbf{\tilde{h}}_1,
\ldots,\mathbf{\tilde{h}}_{L}]$,
$\mathbf{U}=[\mathbf{u}_1,\ldots,\mathbf{u}_L]$, and $\mathbf{1}$ is the
all-ones vector of the required dimension.

\subsubsection{Virtual Uplink Precoder}
Given a fixed $\mathbf{U}$ and $\mathbf{p}$, the optimal decoders
$\mathbf{V}_k$ are the MMSE receivers:
\begin{equation*}
\mathbf{V}_k = \mathbf{J}_k^{-1} \mathbf{H}_k^H \mathbf{U}_k
\sqrt{\mathbf{P}_k}.
\end{equation*}
In this equation, $\mathbf{J}_k = \mathbf{H}_k^H \mathbf{UPU}^H\mathbf{H}_k +
\sigma^2 \mathbf{I}_{N_k}$ is the receive covariance matrix for user $k$.
The optimum virtual uplink precoders are then the normalized columns of
$\mathbf{V}_k$.

\subsubsection{Virtual Uplink Power Allocation}
The power allocation problem on the virtual uplink solves
(\ref{eqn:minpmse}) with a fixed matrix $\mathbf{V}$.  While it is well accepted that the
power allocation subproblem in PMSE minimization (or equivalently, in
sum rate maximization) is non-convex~\cite{SJHU06,BTC06,CTJL07},
recent work~\cite{SSB07} has shown that the optimal power allocation
can be found by formulating the subproblem as a Geometric Programming (GP) problem~\cite{BV04}.  A similar
approach was proposed in~\cite{CTJL07}, where iterations of the the
sum rate maximization problem are solved by local approximations of
the non-convex sum rate function as a GP.  We employ numerical
techniques (SQP) to solve the power allocation subproblem.

In summary, the PMSE minimization algorithm keeps three of four parameters
($\mathbf{U},\mathbf{p},\mathbf{V},\mathbf{q}$) fixed at each step and
obtains the optimal value of the fourth.  Convergence of the overall algorithm to a local minimum is guaranteed since the PMSE objective function is non-increasing at each of the four parameter
update steps.  Termination of the algorithm is determined by the
selection of a convergence threshold $\varepsilon$.

Since the overall minimization problem (\ref{eqn:minpmse}) is not convex, all of the suggested methods are guaranteed to converge only to a local minimum.  Nonetheless, simulations suggest that the locally optimal value of the sum rate is not overly sensitive to selection of an appropriate initialization point.  It is important to ensure that the initial solution allocates power to all $L$ substreams, as the iterative algorithm tends to not allocate power to streams with zero power.  A reasonable initialization is to select random unit-norm precoder vectors in $\mathbf{U}$ and uniform power allocated over all substreams.  A summary of our proposed algorithm can be found in Table~\ref{table:pmse_algorithm}.
%
\begin{table}[!t]
\caption{Iterative PMSE minimization algorithm} 
\centering 
\begin{tabular}{l} 
\hline\hline\\
\textbf{Iteration:}\\
1- \textit{\ \ Downlink Precoder}\\
\qquad $\mathbf{\tilde{U}}_k =\mathbf{J}^{-1}\mathbf{H}_{k}^H\mathbf{V}_{k}
  \sqrt{\mathbf{Q}_{k}}$, \qquad $\mathbf{u}_{kj} = \frac{\mathbf{\tilde{u}}_{kj}}
     {\|\mathbf{\tilde{u}}_{kj}\|_2}$\\ \\
2- \textit{\ \ Downlink Power Allocation via MSE duality}\\
\qquad
$\mathbf{p}=\sigma^{2}(\mathbf{D}^{-1}-\mathbf{\Psi})^{-1}\mathbf{1}$\\
\\
3- \textit{\ \ Virtual Uplink Precoder}\\
\qquad $\mathbf{\tilde{V}}_k = \mathbf{J}^{-1}_{k}\mathbf{H}_{k}^{H}\mathbf{U}_{k}
  \sqrt{\mathbf{P}_{k}}$, \qquad $\mathbf{v}_{kj} = \frac{\mathbf{\tilde{v}}_{kj}}
            {\|\mathbf{\tilde{v}}_{kj}\|_2}$\\ \\
4- \textit{\ \ Virtual Uplink Power Allocation}\\
\qquad $\mathbf{q}=\arg\min_{\mathbf{q}}\prod_{k=1}^{K} \prod_{j=1}^{L_k}
 \epsilon_{kj} $, s.t. $q_{kj} \geq 0$, $\|\mathbf{q}\|_{1} \le P_{\max}$ \\
\\
5- \textit{\ \ Repeat 1--4 until
 $\left[\mathrm{PMSE}_{\mathrm{old}} - \mathrm{PMSE}_{\mathrm{new}}\right]/
 \mathrm{PMSE}_{\mathrm{old}} < \varepsilon$}\\\\
\hline\hline
\end{tabular}
\label{table:pmse_algorithm} 
\end{table}

\section{Numerical Examples\label{section:sims}}
In this section, we present simulation results to illustrate the
performance of the proposed algorithms. In all cases, the fading channel
is modelled as flat and Rayleigh, with i.i.d.\ channel coefficients distributed as $\mathcal{CN}(0,1)$.
The examples use a maximum transmit power of $P_{\max}=1$; SNR is controlled by varying the
receiver noise power $\sigma^2$.  As stated earlier, the transmitter is assumed to have
perfect knowledge of the channel matrix $\mathbf{H}$.

\subsection{Sum Capacity and Achievable Sum Rate}
\begin{figure}[!t]
\centering
\includegraphics[width=\plotwidth]{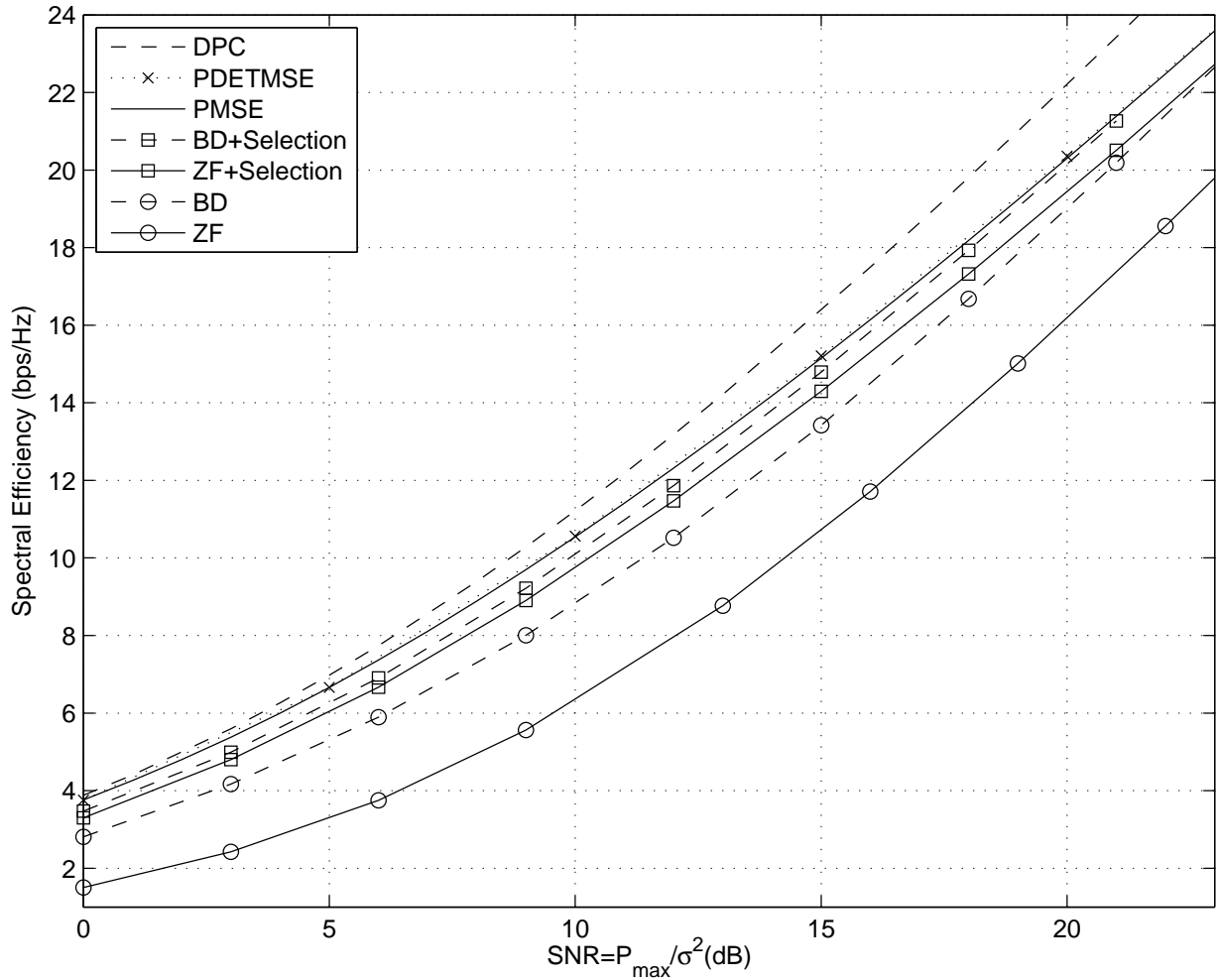}
\caption{Comparing PDetMSE, PMSE, DPC and orthogonalization--based
  methods, $K=2$, $M=4$, $N_k=2$, $L_k=2$}\label{fig:theoretical2Rx}
\end{figure}

We first compare the sum rate achievable using linear precoding to the
information theoretic capacity of the BC.  That is, we consider the spectral
efficiency (measured in bps/Hz) that could be achieved under ideal transmission
by drawing transmit symbols from a Gaussian codebook.
Figure~\ref{fig:theoretical2Rx} illustrates how the proposed schemes 
perform when compared to the sum capacity for the broadcast channel 
(i.e., using dirty paper coding (DPC)~\cite{Costa83}) and to linear 
precoding methods based on channel orthogonalization, i.e., block 
diagonalization (BD) and zero forcing (ZF)~\cite{LJ06}.\footnote{Simulation
results for the DPC, BD, ZF, and NuSVD plots were obtained by using the
\texttt{cvx} optimization package~\cite{CVX,DCP}.}  The convergence
threshold for the PMSE algorithm is set at $\varepsilon=10^{-6}$.  Note that
curves for THP can not be included for comparison, as the modulo and shaping
losses from the DPC sum capacity are fundamentally related to THP's nonlinear
modulation scheme.

%
%
%
%
\begin{figure}[!t]
\centering
\includegraphics[width=\plotwidth]{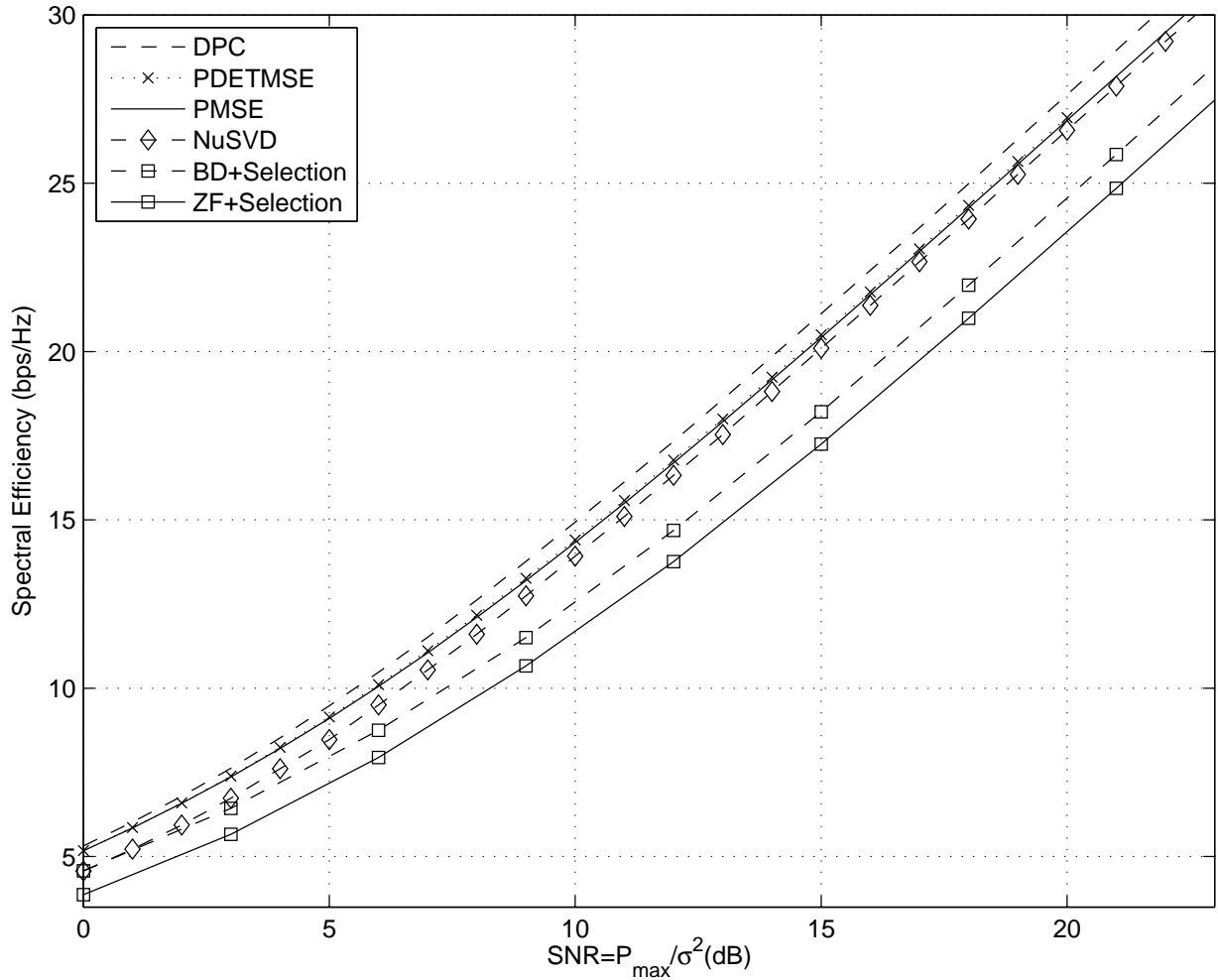}
\caption{Comparing PDetMSE, PMSE, DPC and orthogonalization--based
  methods, $K=2$, $M=4$, $N_k=4$, $L_k=2$}\label{fig:theoretical4Rx}
\end{figure}

The simulations in Fig.~\ref{fig:theoretical2Rx} model a $K=2$ user system with $M=4$ transmit antennas and $N_k=2$ receive antennas per user.  We see a negligible difference in
performance when comparing the PDetMSE algorithm to the PMSE solution.
This is interesting because the relationship between PDetMSE and PMSE
mirrors that of BD and ZF; that is, the PDetMSE can be viewed as the
block-matrix formulation of the PMSE problem. There is, however, a
significant performance difference between BD and ZF.  This result is
also gratifying because it suggests that the marginal gains achieved by
joint processing do not merit the greatly increased computational
complexity; the feasible PMSE solution can be used without a large
penalty in performance.  The PMSE and PDetMSE algorithms do demonstrate
a divergence in performance from the theoretical DPC bound
at higher SNR.  This drop in spectral efficiency may reflect a
fundamental gap between the (optimal) nonlinear DPC capacity and the
rate achievable under linear precoding, but it may also be caused by the algorithms' convergence to local minima due to the non-convexity of the optimization problems.

The PMSE algorithm outperforms the BD and ZF methods over the entire SNR 
range when the orthogonalization-based schemes are forced to use all $N$ receive antennas.  However, this this approach to orthogonalization is suboptimal; the optimal BD and ZF precoders may be found by selecting the best precoder from all $\sum_{k=1}^{\min\left(N,M\right)}\left(\begin{array}{c}N\\k\end{array}\right)$ possible subsets of receive antennas.  At high SNR, the PMSE and PDetMSE precoders perform equivalently to the BD precoder with selection; we have observed that the PMSE and PDetMSE precoders (in conjunction with the MMSE receivers) behave like the BD precoder in orthogonalizing the channel at high SNR.  The biggest gain in performance over 
orthogonalization-based solutions occurs at low to mid-SNR values,  where BD and ZF suffer due to noise enhancement.

Figure~\ref{fig:theoretical4Rx} presents simulation results 
for a similar system as Fig.~\ref{fig:theoretical2Rx}, but with $N_k=4$ receive antennas per user.  
In this system, there are fewer transmit antennas than receive 
antennas ($M<N$), so BD/ZF can not be employed without selection.  We 
include simulation results for BD/ZF with selection, but note 
the large computational complexity required (selecting the best of 
162 candidate precoders).  We compare these results to a generalized 
orthogonalization based approach, referred to as nullspace-directed SVD 
(NuSVD) in~\cite{PWN04}, and observe a large difference in 
performance at high SNR.  This gain in spectral efficiency can be attributed
to NuSVD's ability to use all $N=8$ receive antennas, whereas BD 
and ZF are limited by an antenna constraint.

Once again, Fig.~\ref{fig:theoretical4Rx} illustrates that the 
PMSE/PDetMSE approaches outperform orthogonalization, particularly at low to mid-SNR values.  This improvement in performance comes at the expense of additional complexity.  Even though NuSVD and PMSE/PDetMSE are iterative algorithms, NuSVD requires only one (concave) waterfilling
power allocation after convergence of precoder direction iterations,
whereas the PMSE/PDetMSE minimization methods employ numerical
optimization algorithms (SQP) in each iteration.

\begin{figure}[!t]
\centering
\includegraphics[width=\plotwidth]{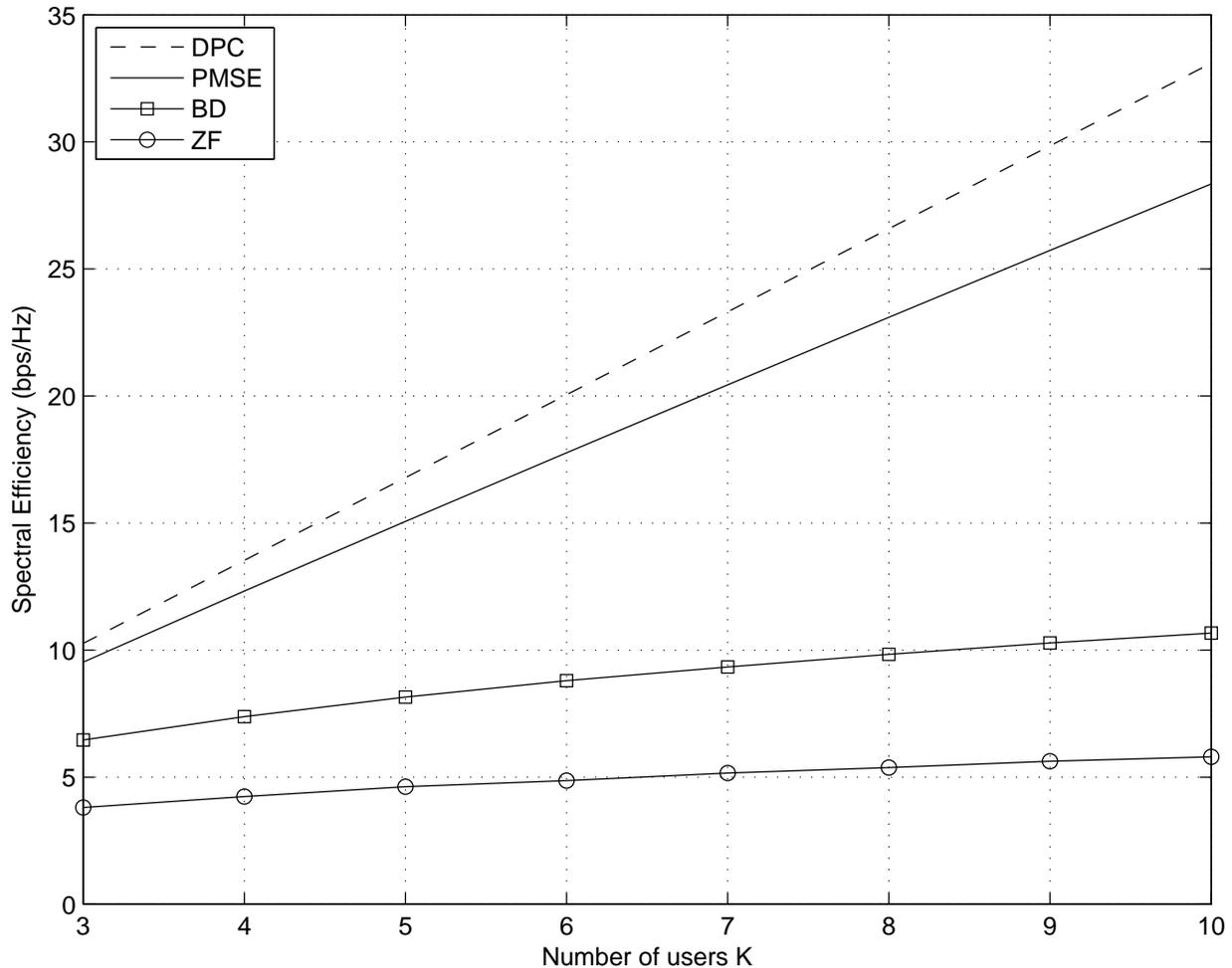}
\caption{Scaling of sum rate with $K$, $M=2K$, $N_k=L_k=2$, $\mathrm{SNR}=5 \mathrm{dB}$}\label{fig:Kplot}
\end{figure}

Figure~\ref{fig:Kplot} shows how the sum throughput scales with the
number of users $K$, for $M=2K$ transmit antennas and $N_k=2$ receive
antennas per user at 5 dB average SNR.
%
%
The number of transmit antennas $M$ is chosen so that BD and ZF can be implemented without selection, as selection-based BD and ZF are exponentially complex with $4^K-1$ possible precoders.  This plot illustrates how the proposed scheme takes advantage of the available degrees of freedom at
the transmitter and provides throughput significantly better than the
orthogonalization based BD and ZF schemes.

\begin{figure}[!t]
\centering
\includegraphics[width=\plotwidth]{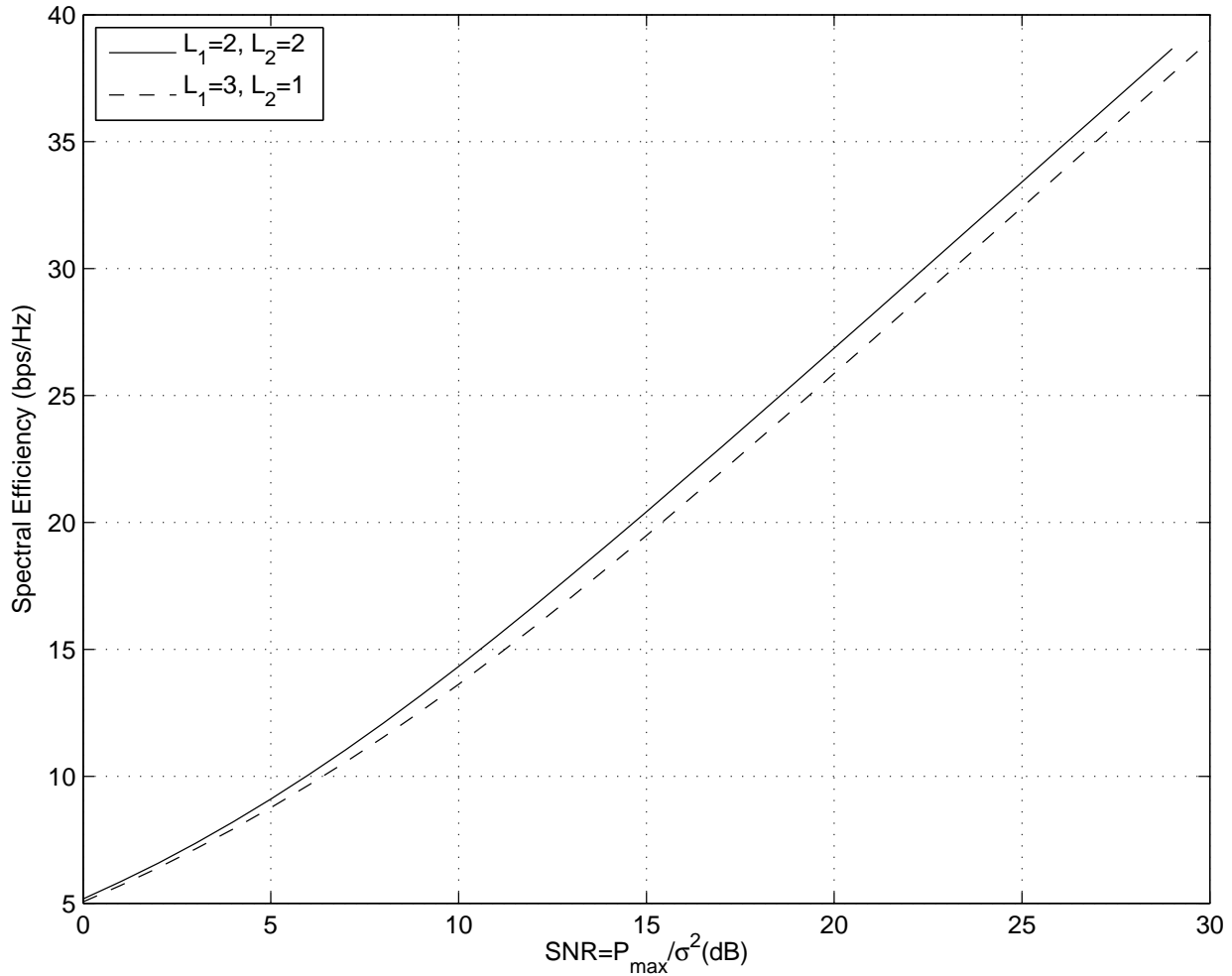}
\caption{Data stream allocation in PMSE optimization, $K=2$, $M=4$, $N_k=4$\label{fig:22vs31}}
\end{figure}

The PMSE and PDetMSE algorithms do not require the explicit selection
of $L_k$; rather, this parameter is determined implicitly by the power
allocation.  However, we can force the PMSE algorithm to allocate a
maximum number of substreams $L_k$ to each user to gain further
insight into its behaviour.  In Fig.~\ref{fig:22vs31}, the number of
streams in the $N_k=4$ system described above is varied from
$L_1=L_2=2$ to $L_1=3$ and $L_2=1$.
%
%
The achievable sum rate in this system decreases in the latter case, as the asymmetric stream
allocation does not correspond to the symmetric (statistically
identical) channel configuration.  In this case, user 2 is restricted
to only a single data stream, and thus can not take full advantage of
good channel realizations.  If the goal is always maximizing the sum
rate, the users should be allocated the maximum number of data streams
in as balanced a manner as possible.  Note however that the PMSE
algorithm can provide unbalanced allocations if desired for other
reasons (e.g., quality of service provisioning).

\subsection{Implementation: Adaptive Modulation\label{sub:adaptivemodulation}}
In contrast to the previous results based on Gaussian codebooks, we
now consider the selection of constellations for modulation to achieve
a maximum throughput for a specified bit error rate (BER) target of
$\beta_{kj}$ on user $k$'s $j^{\mathrm{th}}$ substream.  Since the
PMSE algorithm assumes unit energy symbols, we use $M$-PSK
constellations in our implementation.  Note that the prior assumption of Gaussian noise-plus-interference still holds for a sufficient number of interferers under the central
limit theorem. We propose two approaches for selecting the modulation
scheme for each substream.

\subsubsection{Naive Approach}
This approach selects the largest PSK constellation of $b_{kj}$
bits per stream that satisfies the required BER constraint.  The constraint is satisfied using a
closed form BER approximation~\cite{CG01},
\begin{equation}\label{eqn:mpsk}
\mathrm{BER}_{\mathrm{PSK}}(\gamma) \approx  c_1 \exp \left( \frac{-c_2
\gamma }{2^{c_3 b} - c_4}\right),
\end{equation}
where $M=2^b$ is the size of the PSK constellation.  We apply the least aggressive of the bounds proposed in~\cite{CG01} by using the values $c_1=0.25$,$c_2=8$,$c_3=1.94$, and $c_4=0$.  We note
that this approximation only holds for $b \ge 2$; as such, one can use the exact expression for BPSK:
\begin{equation}\label{eqn:bpsk}
\mathrm{BER}_{\mathrm{BPSK}}(\gamma) = \frac{1}{2}
\mathrm{erfc}\left(\sqrt{\gamma}\right).
\end{equation}

The BPSK expression can be used as a test of feasibility for the
specified BER target; if the resulting BER under BPSK modulation is
higher than $\beta_{kj}$, then we have two options: either declare the
BER target infeasible, or transmit using the lowest modulation depth
available (i.e. BPSK).  In this work, we have elected to transmit using
BPSK whenever the PMSE stage has allocated power to the data stream.

\subsubsection{Probabilistic Approach}
The naive approach is quite conservative in that there may be a large gap
between the BER requirement and $\mathrm{BER}_{b_{kj}}$, the BER achieved for each channel realization.  We suggest a \textit{probabilistic} bit allocation scheme that switches between
$b_{kj}$ bits (as determined by the naive approach) and $b_{kj}+1$ bits with
probability
$p_{kj} = \left[\beta_{kj} - \mathrm{BER}_{b_{kj}}\right]/
\left[\mathrm{BER}_{b_{kj}+1} - \mathrm{BER}_{b_{kj}}\right]$.  This
modulation strategy may not be appropriate for systems requiring
instantaneous satisfaction of BER constraints; however, the
probabilistic method will still achieve the desired BER in the
long-term over multiple channel realizations.

\begin{figure}[!t]
\centering
\includegraphics[width=\plotwidth]{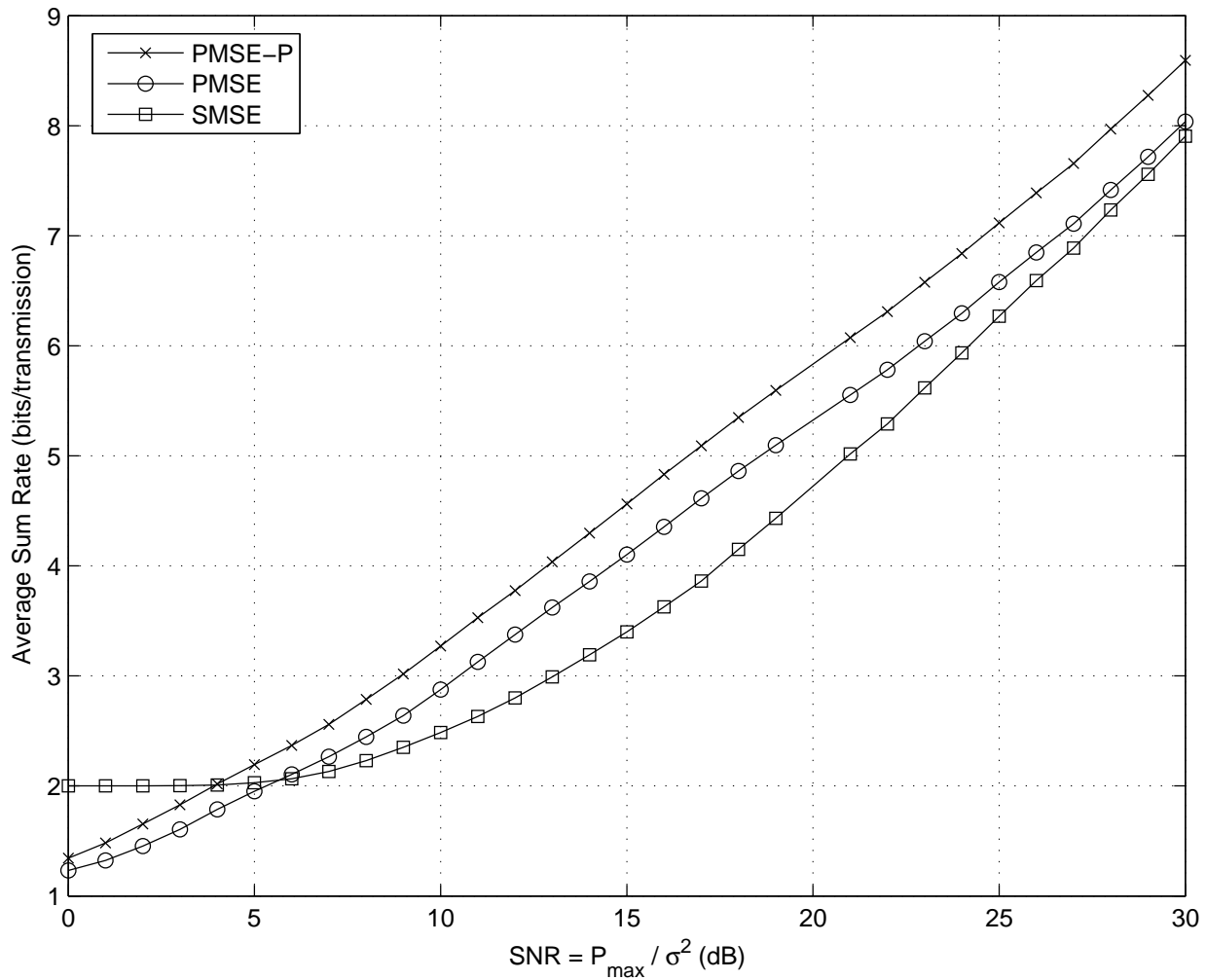}
\caption{Sum rate vs. SNR for user 1 with M-PSK modulation, $K=2$,
  $M=4$, $N_k=L_k=2$}\label{fig:sumrate}
\end{figure}

Figure~\ref{fig:sumrate} shows the sum rate achieved in the same system
configuration as in Fig.~\ref{fig:theoretical2Rx} ($K=2$, $M=4$, $N_k=2$) under the $M$-PSK modulation scheme described above.
%
%
The simulations use two data streams per user and a target bit error rate of $\beta_{kj} = 10^{-2}$; $5000$ data and noise realizations are used for each channel realization.  The plot illustrates the average number of bits per transmission for user 1; due to symmetry, the corresponding plot for user 2 is identical.  Note that in contrast to the previous results based on Gaussian coding using spectral efficiency, the sum rate in Fig.~\ref{fig:sumrate} is the average number of bits transmitted per realization using symbols from a PSK constellation.

In Fig.~\ref{fig:sumrate}, we consider using the PSK modulation scheme for the PMSE precoder and the SMSE precoder designed in~\cite{KTA06}.  Examination of this plot reveals that using the PMSE
criterion is justified at practical SNR values with improvements of
approximately one bit per transmission near 15 dB. Furthermore, using the
probabilistic modulation scheme (designated ``PMSE-P'') yields an
additional improvement of more than half a bit per transmission across
all SNR values.

\begin{figure}[!t]
\centering
\includegraphics[width=\plotwidth]{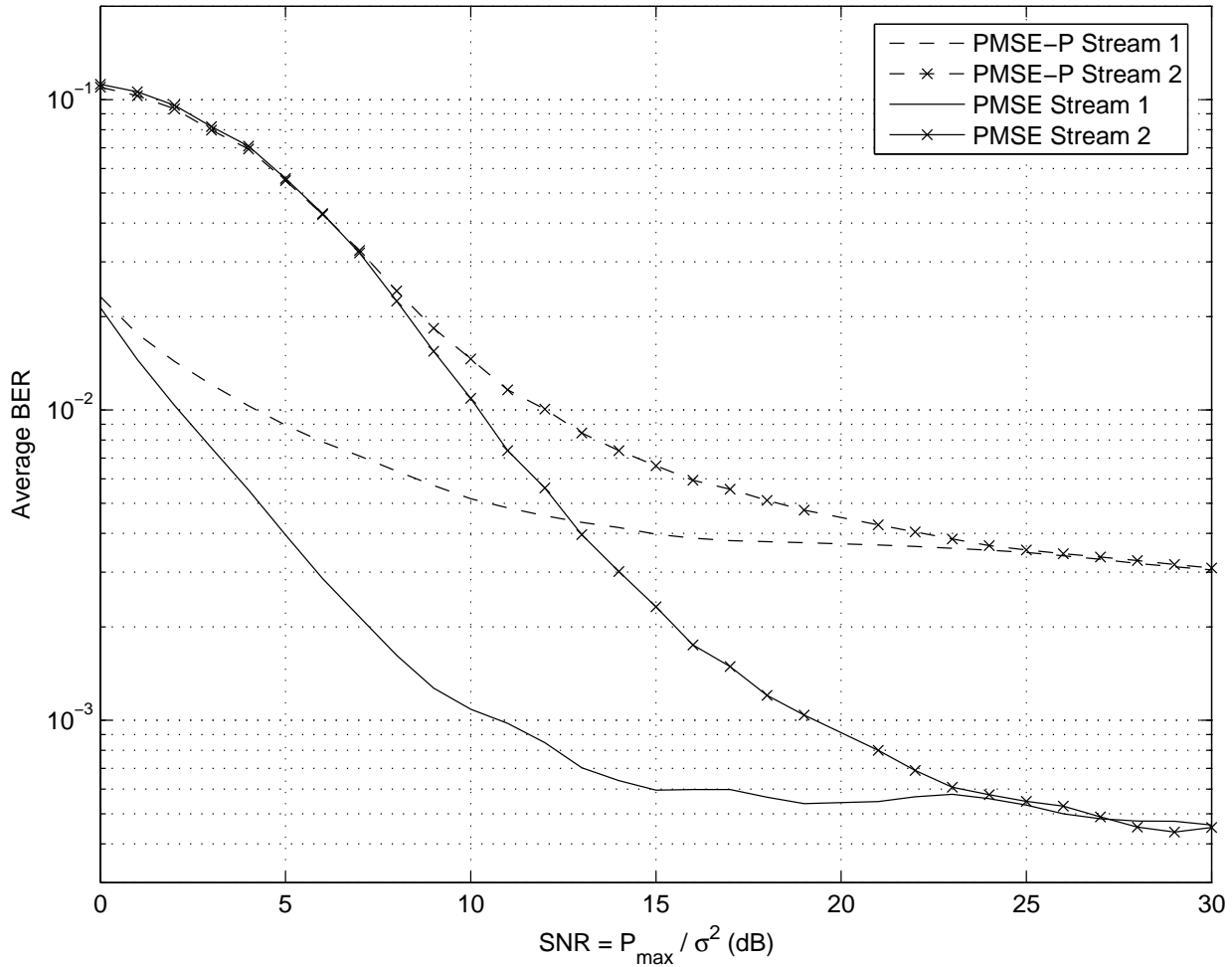}
\caption{BER vs. SNR for user 1 with M-PSK modulation, $K=2$, $M=4$, $N_k=L_k=2$}\label{fig:ber}
\end{figure}

In Fig.~\ref{fig:ber}, we plot average BER versus SNR for the same system
configuration as in Fig.~\ref{fig:sumrate}.
%
%
This plot illustrates how the naive bit allocation algorithm attempts to achieve the target BER of $10^{-2}$ for all data streams under PMSE, but also overshoots the target, converging to
a BER of approximately $5 \times 10^{-4}$.  This can be attributed to the
looseness of the BER bound mentioned above.  In contrast, the probabilistic rate
allocation algorithm not only increases the rate, as shown in
Fig.~\ref{fig:sumrate}, but also converges to a BER that is much closer to the
desired target BER. The remaining gap between the actual BER achieved and the
target BER can again be attributed to looseness in the approximations
of (\ref{eqn:mpsk}) and (\ref{eqn:bpsk}).

\section{Conclusions}\label{section:conc}
In this paper, we have considered the problem of designing a linear precoder to maximize
sum throughput in the multiuser MIMO downlink under a sum power
constraint. We have compared the maximum achievable sum rate
performance of linear precoding schemes to the sum capacity in the
general MIMO downlink, without imposing constraints on the number of
users, base station antennas, or mobile antennas.  The problem was
reformulated in terms of MSE based expressions, and the joint
processing solution based on PDetMSE minimization was shown to be
theoretically optimal, but computationally infeasible.  A suboptimal
framework based on scalar (per-stream) processing was then proposed,
and an implementation was provided based on PMSE minimization and
employing a known uplink-downlink duality of MSEs.  We evaluated
the performance of these schemes in the context of orthogonalizing
approaches, which suffer from noise enhancement, and have shown that the MSE based optimization schemes
are able to achieve significant performance improvements.
Furthermore, we have demonstrated that negligible performance losses
occur when using the suboptimal PMSE criterion in comparison to the
optimum PDetMSE criterion.
%
\bibliographystyle{IEEEtran}

\end{document}